\documentclass[11pt,twoside]{article}
\usepackage{asp2014}

\aspSuppressVolSlug
\resetcounters

\begin{document}

\title{Availability of Hyperlinked Resources in Astrophysics Papers}

\author{P. Wesley~Ryan,$^1$ Alice Allen,$^{1,2}$ and Peter Teuben$^2$}
\affil{$^1$Astrophysics Source Code Library; \email{wes@ascl.net}}
\affil{$^2$Astronomy Department, University of Maryland, College Park}

\paperauthor{P. Wesley~Ryan}{wes@ascl.net}{ORCID}{Astrophysics Source Code Library}{~}{~}{~}{~}{~}
\paperauthor{Alice~Allen}{}{ORCID}{Astrophysics Source Code Library}{~}{~}{~}{~}{~}
\paperauthor{Teuben~Peter}{teuben@astro.umd.edu}{0000-0003-1774-3436}{Astronomy Department}{University of Maryland}{College Park}{MD}
{20742}{USA}


\begin{abstract}

Astrophysics papers often rely on software which may or may not be available, and URLs are often used as proxy citations for software and data. We extracted all URLs from two journals' 2015 research articles, removed those from certain long-term reliable domains, and tested the remainder to determine what percentage of these URLs were accessible in October 2018. 
  
\end{abstract}


\section{Introduction}

Astrophysics, like most disciplines, relies on software and repositories of data for its research. Software source codes and raw data are often too large and complex to share within the papers in which they are used. Although standards for citing software, such as the IDs assigned by the Astrophysics Source Code Library (ASCL)\footnote{\url{https://ascl.net}}, have recently been developed and are increasingly used, they have yet to reach universal adoption. As a result, information relevant to a paper may be accessible from that paper only via a URL. \citet{howisonbullard2015} found that 5\% of software mentions in a sample of biology papers referenced the software only by a link in the text. (In comparison, 31\% used only an in-text mention of the name of the software.) Even when software is properly cited, it may be available only as a binary or web tool, or not available at all.

Although some pre-WWW hypertext system designs provided for guaranteed resource persistence, the Web itself does not; as a result, URLs may fail to resolve to the proper resource. Websites are reorganized, graduate students move on, and professors retire. In each case, links may break, including links contained in published papers. 

Resource accessibility is important to research transparency, repeatability, and reproducibility. If the code used to compute a result is not available, it cannot be audited for bugs; if the data upon which a computation is carried out is inaccessible, it cannot be reanalyzed. Because of the importance of resource accessibility, and because these resources may be referred to with hyperlinks, we have undertaken a research project dedicated to studying the availability of hyperlinked resources over time.

Similar studies have been carried out in other fields. \citet{manguletal2018} found that 24\% of URLs in a large sample of biomedical papers published from 2000 to 2017 were broken, and 4\% more were unreachable due to connection timeouts.

\section{Context and methods}

The present paper is a follow-up to a 2018 study conducted by the authors (\citet{allenteubenryan2018}). As part of that study, we extracted the HTTP(S) and FTP hyperlinks from all papers published in \emph{Astronomy \& Astrophysics} (\emph{A\&A}) in 2015, excluded links to nine commonly-referenced `infrastructure' sites that we knew to be available, tested the rest for availability, and assigned them to one of three categories: consistently available, consistently unavailable, and inconsistently available. For HTTP(S) links, we used an automated checker derived from one in use at the Astrophysics Source Code Library, and defined consistently available links as those which returned the \texttt{200 OK} status code every time we tested them; consistently unavailable links as those which returned other status codes, contained domain names which could not be resolved, or had other errors\footnote{Failed domain lookups and other errors that do not correspond to HTTP status codes were assigned codes with negative numbers.}; and inconsistently available links as those which sometimes returned \texttt{200 OK} status codes and sometimes did not. Links that were inconsistent in the status codes they returned but never returned a \texttt{200 OK} status code were categorized as consistently unavailable.

Due to the small number of FTP links contained in our dataset and the limitations of the ASCL link checker, we opted in our original research to check these links by hand, and assign them to our categories by whether or not they resolved to a resource.

In our follow-up research, we tested the same HTTP(S) links one year after our initial checks (which were carried out in September and October 2017), and extracted and tested the HTTP(S) links in the papers published in the \emph{Astrophysical Journal} (\emph{ApJ}) in 2015, using the \texttt{LExTeS} package \citep{lextes} that we developed for the initial paper, with minor improvements.\footnote{The original version of \texttt{LExTeS} was written in Python 2 and designed to be run on Linux. Since the present author now has Python 3 and Windows, the software was ported from Python 2 to Python 3 and certain Linux-specific idioms, such as the expectation of Unix-style shell glob expansion, were replaced with more portable ones. In addition, the now-deprecated library \texttt{pyPdf}, used for extracting hyperlinks from PDFs, was replaced with \texttt{PyPdf2}.} In this paper, we consider only HTTP(S) links; since the \emph{A\&A} dataset contains only 30 FTP links and the \emph{ApJ} dataset contains only 45, the effect of discarding them is negligible.


\section{Results}

Table 1 shows the percentage of links by availability category. Of the 2,528 HTTP(S) links in the \emph{A\&A} dataset, 2,176 were consistently available, 322 were consistently unavailable (4 of which were consistently unavailable but for inconsistent reasons, so are not listed in Table 2), and 30 were inconsistently available. Of the 3,141 HTTP(S) links in the \emph{ApJ} dataset, 2,626 were consistently available, 460 were consistently unavailable (6 of which were consistently unavailable but for inconsistent reasons, so are not listed in Table 2), and 55 were inconsistently available.

\begin{table}[!ht]
\caption{Percentage of links by category}
\smallskip
\begin{center}
{\small
\begin{tabular}{lccc}  
\tableline
\noalign{\smallskip}
~ & Up & Down & Inconsistent\\
\noalign{\smallskip}
\tableline
\noalign{\smallskip}
\emph{A\&A} (Sep./Oct.~2017) & 86.8\% & 10.6\% & 2.6\% \\
\emph{A\&A} (Oct.~2018) & 86.1\% & 12.7\% & 1.2\% \\
\emph{ApJ} (Oct.~2018) & 83.6\% & 14.6\% & 1.8\% \\
\tableline\
\end{tabular}
}
\end{center}
\end{table}

\begin{table}[!ht]
\caption{Consistently unavailable links that always returned the same error code}
\smallskip
\begin{center}
{\small
\begin{tabular}{lcc}
\tableline
\noalign{\smallskip}
Error code & \emph{A\&A} (Oct.~2018) & \emph{ApJ} (Oct.~2018) \\
\noalign{\smallskip}
\tableline
\noalign{\smallskip}
-7 Timeout error & 0 & 1 \\
-6 Connection reset & 0 & 2 \\
-5 Value error & 0 & 9 \\
-4 Bad status line & 1 & 0 \\
-3 Socket error & 2 & 0 \\
\tableline
-2 SSL certificate error & 6 & 0 \\
-1 Lookup failed & 97 & 172 \\
302 Found & 0 & 0 \\
400 Bad request & 0 & 2 \\
401 Unauthorized & 3 & 7 \\
\tableline
403 Forbidden & 34 & 32 \\
404 Not found & 167 & 220 \\
500 Internal server error & 5 & 5 \\
502 Bad gateway & 1 & 1 \\
503 Service unavailable & 2 & 3 \\
\tableline\
\end{tabular}
}
\end{center}
\end{table}

\section{Conclusions and Future Work}

The rate of link decay seems not to be constant over time. Although a small number of the links in the \emph{A\&A} dataset that were categorized as consistently unavailable in 2017 recorded some successful checks in 2018, the overall percentage of broken links shows a lower per-year decrease from 2017 to 2018 than from 2015 to 2017.  

We plan to test our current datasets of URLs periodically and build additional datasets of links from other years and other journals. Our goal is to track link health and the degree of reliance on possibly ephemeral methods of referencing code and data, and investigate whether recent and continuing changes in citation methods improve the overall availability of these resources going forward.

\bibliography{P13-15}

\end{document}